# Damping of optomechanical disks resonators vibrating in air

D. Parrain[1], C. Baker[1], T. Verdier[1], P. Senellart[2], A. Lemaitre[2], S. Ducci[1], G. Leo[1], I. Favero[1 *]

[1] *Université Paris Diderot, Sorbonne Paris Cité, Laboratoire Matériaux et Phénomènes Quantiques,*

*CNRS-UMR 7162, 10 rue Alice Domon et Léonie Duquet, 75013 Paris, France*

[2] *Laboratoire de Photonique et Nanostructures, CNRS, Route de Nozay, 91460 Marcoussis, France*

We report on miniature GaAs disk optomechanical resonators vibrating in air in the radiofrequency range. The flexural modes of the disks are studied by scanning electron microscopy and optical interferometry, and correctly modeled with the elasticity theory for annular plates. The mechanical damping is systematically measured, and confronted with original analytical models for air damping. Formulas are derived that correctly reproduce both the mechanical modes and the damping behavior, and can serve as design tools for optomechanical applications in fluidic environment.

Optomechanics studies the coupling of light to mechanical motion, with applications in the quantum-optical control of mechanical systems, in optical-mechanical sensing or in metrology[1-3]. While many recent developments in optomechanics have striven for reduced optical and mechanical dissipation to unravel quantum phenomena[4-5], more dissipative regimes are also of interest. For example, the operation of optomechanical systems in air or in a liquid, where their dissipation is increased, calls for attention if these systems are to serve as generic force or mass sensors[6]. In complex fluids, rheological studies employing miniature mechanical resonators[7] can also benefit from the un-surpassed sensitivity of optomechanical systems, provided that their interaction with the fluid is controlled. Miniature whispering-gallery resonators in the form of silica toroids[8] and semiconductor disks[9-11] offer an ultra-large optomechanical coupling, which makes them systems of choice for optomechanics experiments. However, the resonators lack simple models to describe their mechanical properties: their mechanical modes are generally computed by numerical methods and studies of their mechanical damping in a fluid are scarce.

In this letter, we advance towards an understanding of the mechanics of optomechanical disk structures operated in a fluid. We investigate the mechanical behavior of Gallium-Arsenide (GaAs) disk resonators vibrating in air, by performing Scanning Electron Microscope (SEM) and optical fiber interferometric detection of their mechanical motion. We show that the elasticity theory of annular plates describes well the disk's mechanical modes and allows deriving effective analytical expressions for their eigenfrequencies. We develop a simple model for air damping of the disk's motion, where Stokes spheres are attached to a vibrating disk. The model is shown to satisfactorily reproduce experimental results on a large set of measured resonators. In the investigated situation, mechanical damping of GaAs disks appears to be dominated by squeeze-film and air damping contributions.

We employ a semi-insulating GaAs substrate on which we grow an epitaxial GaAs 500 nm buffer layer, a 1.5 µm $Al_{0.8}Ga_{0.2}As$ layer and finally a 200 nm GaAs top layer. GaAs-based resonators have shown a low-level of mechanical dissipation in previous work[9,10,12,13]. Here, disks of different diameter (10 to 50 µm) are defined by e-beam lithography with a negative resist, and then fabricated in a two-step wet etching process: first a non-selective etch of GaAs/AlGaAs and then a selective under-etch of the AlGaAs sacrificial layer[14]. The SEM picture of a GaAs disk suspended by this process over an AlGaAs pedestal is shown in Fig. 1a, together with its schematic vertical section in Fig. 1b. These disks possess an important variety of mechanical modes: here we will focus on flexural modes. We first test the theory of elastic annular plates[15] as a means to obtain an analytical description of these flexural modes, taking *a* (*b*) (see Fig.1b) as the annular plate external (internal) radius. Assuming perfect rotational invariance, the modes for the disk flexural motion are represented by an out-of-plane vibration profile $w_{P,M}(r,\theta,t)=W_{P,M}(r,\theta)\cos(\omega_{P,M}t)$ where P and M are the radial and azimuthal numbers. The spatial mode profile is $W_{P,M}(r,\theta)=\cos(M\theta)[A_M J_M(k_{P,M}r)+B_M Y_M(k_{P,M}r)+C_M I_M(k_{P,M}r)+D_M K_M(k_{P,M}r)]$ with $J_M$ and $Y_M$ ($I_M$ and $K_M$) the M-order (modified) Bessel functions of first and second kind, respectively. The wave vector k relates to the angular mechanical frequency by $k^4=12\rho\omega^2(1-\nu^2)/(Et^2)$ with $\rho$ the material density, $\nu$ the Poisson ratio, E the Young's modulus and t the disk thickness. The mechanical frequency and constants A,B,C and D are found by imposing clamping conditions for the disk on the pedestal (r=*b*) and free vibration conditions at the disk end (r=*a*). Fig.1c shows the motion profile of a (P=0,M=1) flexural mode for a clamped annular plate described by this approach.

In our experiments, we excite the mechanical modes of the disks by ultrasonic piezoelectric means: the sample is glued on a piezo-plate driven by an AC voltage. Sweeping the frequency while monitoring the disk motion reveals the mechanical resonances of the system. The disk motion is first analyzed in the SEM chamber (vacuum $10^{-5}$ mbar). Fig. 1d shows for example the resonant excitation of the lowest frequency disk flexural mode, identified as being the (P=0,M=1) mode predicted by elasticity theory. This in-situ SEM excitation technique offers the advantage of directly imaging the vibration profile[16-17] but obtaining a detectable motion requires driving the mechanical resonator with a large piezo power, where the disk mechanical response becomes non-linear. To work in a linear regime, the disk motion is excited at low power and optically measured with a fiber interferometric technique[18-20]. To this purpose, a single-mode optical fiber for $\lambda=1.55\mu m$ is cleaved with a straight angle and positioned over the disk top surface at a distance shorter than the Rayleigh distance. Monochromatic laser light is injected into the fiber, exits the fiber output and circulates in the multiple-cavity system formed by the sample substrate, the disk and the fiber end. This light is modulated by the disk flexural vibration, collected back by the fiber and sent on a photodetector to allow vibration detection. Transfer Matrix simulations show that interferences between the disk and the substrate dominate the behavior of the multiple-cavity system, and that the cavity formed between the disk and the fiber end plays a second-order role in the interferometer. The schematic experimental set-up is shown in Fig 2a. A typical mechanical spectrum obtained with this fiber-technique is shown in Fig 2b, when sweeping the piezo-actuation frequency with the output of a network analyzer and collecting the photodetector signal into its input port. The shape of the obtained mechanical resonance is correctly fitted by the damped harmonic oscillator model, which predicts a motional amplitude response $|x(\omega)|=[1/(\omega_m^2-\omega^2)+(\Gamma\omega)^2]^{1/2}$ where $\omega_m$ is the mechanical angular eigenfrequency and $\Gamma$ the mechanical damping factor, with the mechanical Q factor given by $Q=\omega_m/\Gamma$. The symmetric shape of the resonance indicates a linear mechanical response, as expected for an excitation power 20 dB lower than in the SEM experiment. The linear behavior allows using the position and width of the resonance to analyze the disk elastic and dissipative mechanical properties. Let us first focus on measured mechanical frequencies $f_m=\omega_m/2\pi$. In the theory of annular plates presented above, we cast the mechanical frequency $f_{P,M}$ of a flexural (P,M) mode in the simple form:

$$f_{P,M} = \frac{\lambda_{P,M}^2}{2\pi}\sqrt{\frac{E}{12\rho(1-v^2)}}\frac{t}{a^2} \qquad (1)$$

with a dimensionless parameter $\lambda_{P,M}^2$ defined by $\lambda_{P,M}^2=k_{P,M}^2 a^2$, which depends on (P,M) and can be shown to depend on the disk inner (*b*) and outer (*a*) radius only through their ratio *b/a*. Hereafter, we will focus on (P=0,M=0) flexural modes, as they possess the highest symmetry. Fig 2c plots the evolution of $\lambda_{0,0}^2$ as a function of *b/a*. Each point in this curve is obtained by finding the root of the implicit equation obtained from mechanical boundary conditions. This equation is invariant for geometry changes that let *b/a* invariant. Along with Eq.1, Fig 2c allows evaluating by hand the mechanical frequency for the (P=0,M=0) flexure of an arbitrary elastic disk. Furthermore, let us note that the quantity $\lambda_{0,0}^2(a-b)^2/a^2$, which only depends on the parameter b/a, barely varies (± 10%) when b/a varies from 0 to 0.8 (not shown here). This implies that at a first level of description, the dependence of $f_{0,0}$ is well captured by a simple dependence in $1/(a-b)^2$, reminiscent of the usual $1/l^2$ dependence for the flexural motion of an elastic lever of length l [21]. This is the picture that we will employ below.

Fig 3a shows the $f_{0,0}$ frequency measured in the fiber interferometer on a set of 24 disks. The dimensions of the disks are measured by combining optical microscope and SEM inspection, resulting in an estimated error bar of ±200 nm in the outer and inner diameter of the rotation-invariant representation of Fig1b. The outer diameter $a$ varies between 12 and 23 µm, the inner radius $b$ between 5 and 16 µm, the under-etch distance ($a$-$b$) between 6 and 7 µm and the ratio $b/a$ between 0.44 and 0.72. The measured frequency varies between 2.3 and 3.2 MHz, and decreases for increasing values of ($a$-$b$). The two solid lines in Fig 3a are obtained from the above effective elasticity approach where the frequency scales with the inverse of ($a$-$b$)$^2$, using the two bounding values of $\lambda_{0,0}^2(a-b)^2/a^2$ corresponding to ($b/a$)=0.44 and 0.72. GaAs is treated as an isotropic elastic material with E=85.9 GPa, ν=0.31, ρ=5316 kg/m$^3$, and we use a disk thickness t of 200 nm inferred from in-situ control during the epitaxial growth. The agreement with data shows that this effective analytical theory correctly evaluates the frequency of a GaAs disk flexural mode, and reproduces its dependence on geometric dimensions. For a finer level of agreement, the full dependence of the mechanical frequency on both $a$ and $b$ can of course be employed, or Finite Element Method (FEM) simulations taking the disk pedestal into account[5]. However, the effective analytical approach tested positively here provides a perfect tool to understand the disk elastic properties, and a starting point to model its interaction with its air environment, as we will see below.

Figure 3b reports the mechanical Q factor for the (P=0,M=0) mode of the disks measured in the interferometer. The typical Q is of a few tens and decreases for increasing under-etch ($a$-$b$). To interpret these results, clamping losses are first evaluated numerically in FEM simulations, by adapting a Perfectly-Matched-Layer approach to the problem of acoustic radiation of the disk into the support[22]. For the disks and modes investigated here, these numerical simulations lead to clamping Q factors typically between $10^4$ and $10^5$. This is a negligible contribution in the present experiments; hence we consider next the dissipative effects associated to the presence of air around the disks. These effects are twofold: air damping of the flexural motion, which would be present even without substrate, and squeeze film damping, due to the presence of the substrate directly under the disk. When the Reynolds number for the airflow is low, a model for air damping of thin cantilevers consists in attaching virtual Stokes spheres of radius R along the lever and summing the spheres contributions[23-25]. Here we develop a model for the disk flexural motion, considering the dynamic viscous coefficient for a sphere oscillating at angular frequency $\omega_m$[26]: $\beta=6\pi\eta R(1+R/\delta)$. In this expression $\delta$ is the boundary layer thickness given by $\sqrt{(2\eta/\rho_a\omega_m)}$. With $\eta=1.8\ 10^{-5}$ Pa.s and $\rho_a=1.2$ kg.m$^{-3}$ for the air dynamic viscosity and density, $\delta$ is of the order of the lateral dimensions of the disks investigated here. If the sphere radius R is commensurable with the disk radius, $R/\delta$ cannot be neglected, in contrast to the case of kHz cantilevers. The energy $U_d$ dissipated per vibration cycle is obtained by integrating the work of the dissipative force $-\beta dw_{0,0}/dt$ on the disk surface and averaging over a cycle leading $U_d=6\pi\eta\omega_m(1/R+1/\delta)\iint_{disk} rW^2_{0,0}(r,\theta)\,drd\theta$, where $\omega_m$ is the mechanical angular frequency. This energy is compared to the mechanical energy stored in the mechanical mode $U_s=\omega^2_m\rho t/2 \iint_{disk} rW^2_{0,0}(r,\theta)\,drd\theta$ and we finally obtain the quality factor $Q=2\pi U_s/U_d$:

$$Q_{air} = \frac{\rho t \omega_m}{6\eta} \frac{1}{\frac{1}{R}+\frac{1}{\delta}}$$

Let us now take R=(*a-b*)/2 in the spirit of the cantilever case, where satisfactory results are obtained with a sphere diameter equal to the cantilever width[25]. In this case, our derived $Q_{air}$ decreases for increasing (*a-b*) as observed in the measurements, but varies between 100 and 170 for the disks investigated here, above measured values. This means that, according to our model, air damping indeed contributes to the dissipation of the disks flexural motion, but only accounts for part of it. We consider next the presence of the nearby substrate at a distance h=1.5 µm below the disk, which produces squeeze-film effects. In a simplified approach, we approximate the (P=0,M=0) flexural mode of the disk by the motion of a rigid annular plate in a direction normal to the substrate. An analytical treatment of squeeze-film mechanisms, in the relevant limit of small Reynolds number and isothermal squeezed gas[27], leads us to a Q factor $Q_{squeeze}=\omega_m/\Gamma_{squeeze}$ where:

$$\Gamma_{squeeze} = \frac{3\eta a^2}{2t\rho h^3}\left[1+\left(\frac{b}{a}\right)^2+\frac{1-(b/a)^2}{\ln(b/a)}\right]$$

To obtain satisfactory quantitative results with this formula, an effective over-length for the plate lateral dimension (*a-b*) must be considered to account for border effects. For the small Knudsen number in our study, we adopt an approximate effective over-length of 8h/3π inferred from previous works on moving plates[27-28]. This leads to $Q_{squeeze}$ values that decrease for increasing (*a-b*) and that are about 20% above the measured Q values. This quantitative discrepancy vanishes if we now sum the two contributions for both air damping and squeeze-film damping. To this purpose, we first divide the air damping contribution by a factor two, in order to only account for air damping on the upper side of the disk. Indeed the damping on lower side of the disk is fully included in the squeeze-film contribution. Finally, the two dashed lines in Fig 3b correspond to $1/Q=1/2Q_{air}+1/Q_{squeeze}$ for the two bounding curves for $\omega_m$ in Fig 3a. These dashed lines satisfactorily bound the observed Qs, showing the validity of the employed effective models for air and squeeze-film damping in our experimental situation. An exhaustive systematic study as a function of the disk and surrounding gas parameters is beyond the scope of the present article, but will allow understanding more precisely the limits of validity of these models.

In summary, we study experimentally the elastic and dissipative mechanical behavior of miniature GaAs disks vibrating in air. We derive effective analytical models that satisfactorily explain the observed vibrational mode shape, the measured mechanical frequency and the observed damping. The latter appears to originate from both squeeze-film damping and damping by the surrounding air. As compared to numerical simulations, this analytical modeling offers a direct insight into the parameters that need to be controlled when using optomechanical disk resonators in a fluid. On a more general level, optomechanical disk resonators could become a model system for investigating fluid-structure interactions, thanks to their high geometric symmetry and exquisite optomechanical response to external forces.

This work was supported by CNano Ile de France and the French Agence Nationale de la Recherche.

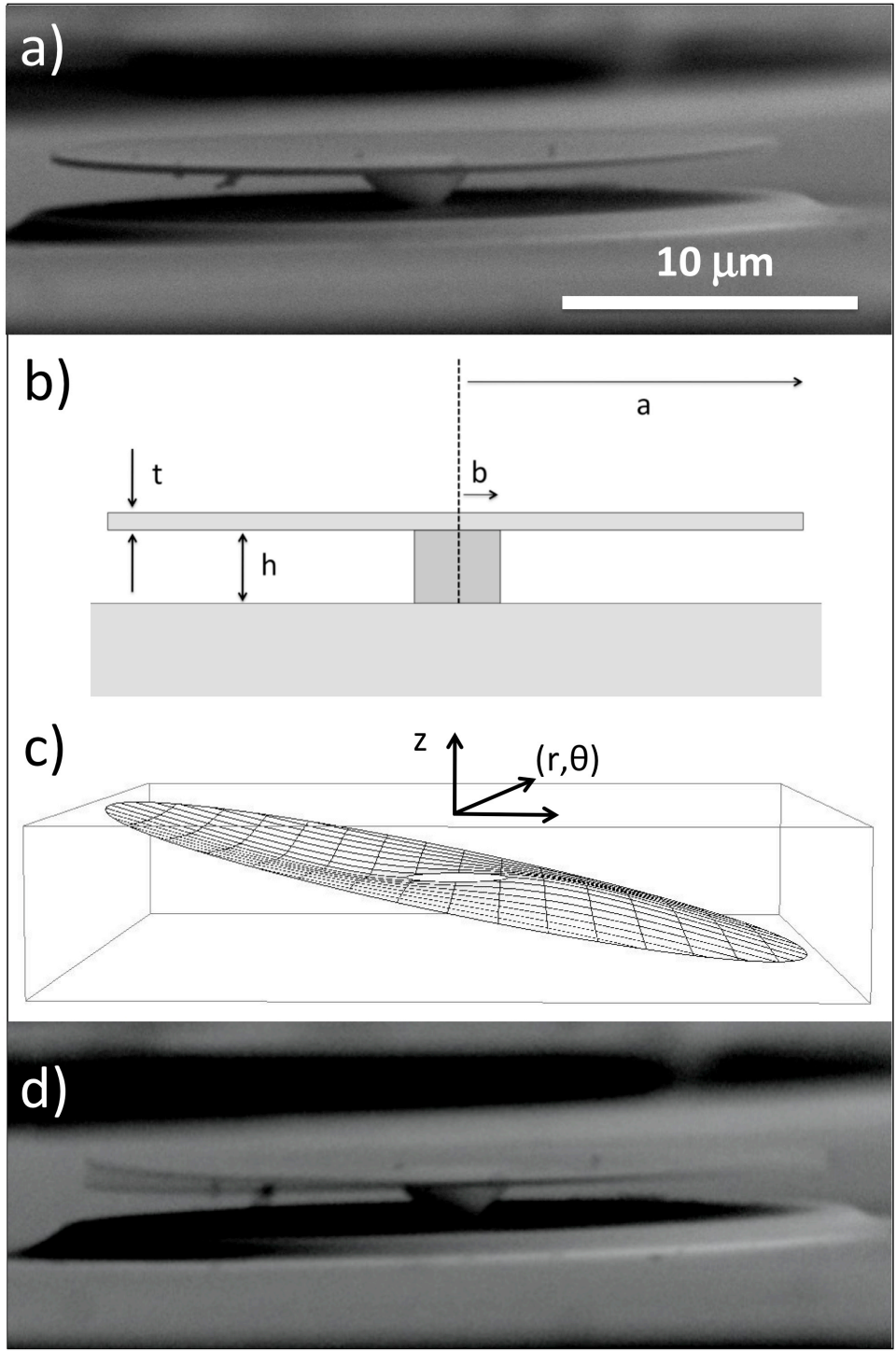

FIG. 1. Flexural motion of micron-sized GaAs disk resonators. a) SEM side-view of a GaAs disk suspended over an AlGaAs pedestal. b) Schematics of the disk geometry with relevant dimensions. c) Vibration profile of a (P=0,M=1) flexural mode represented in cylindrical coordinates. d) SEM side-view of the GaAs disk shown in a), vibrating resonantly on its (P=0,M=1) flexural mode.

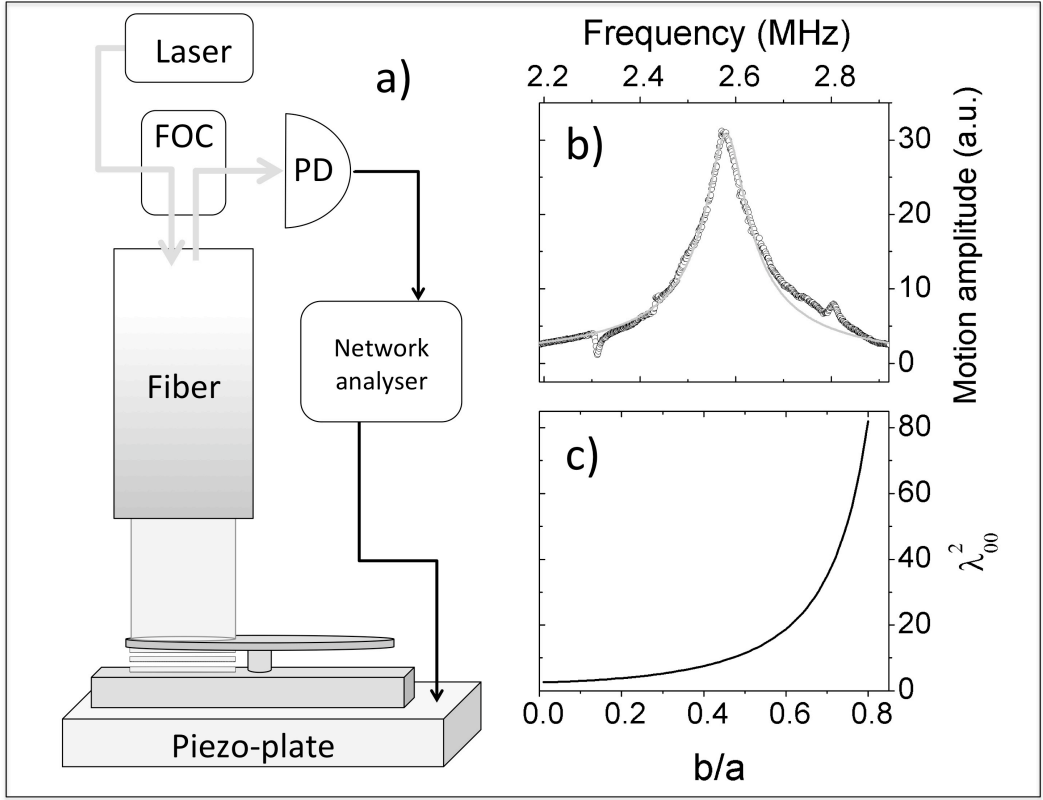

FIG. 2. a) Schematics of the fiber-interferometer set-up. FOC stands for Fiber Optical Circulator, PD for PhotoDetector. b) Vibrational spectrum on a GaAs disk flexural mode. This spectrum is obtained by dividing the raw motional spectrum, obtained when placing the fiber above the disk periphery, by a reference spectrum for piezo-motion obtained with the fiber placed above the rigid disk pedestal. c) The calculated effective parameter $\lambda_{0,0}^2$ entering the effective formula for the (P=0,M=0) mode frequency, and its dependence on b/a.

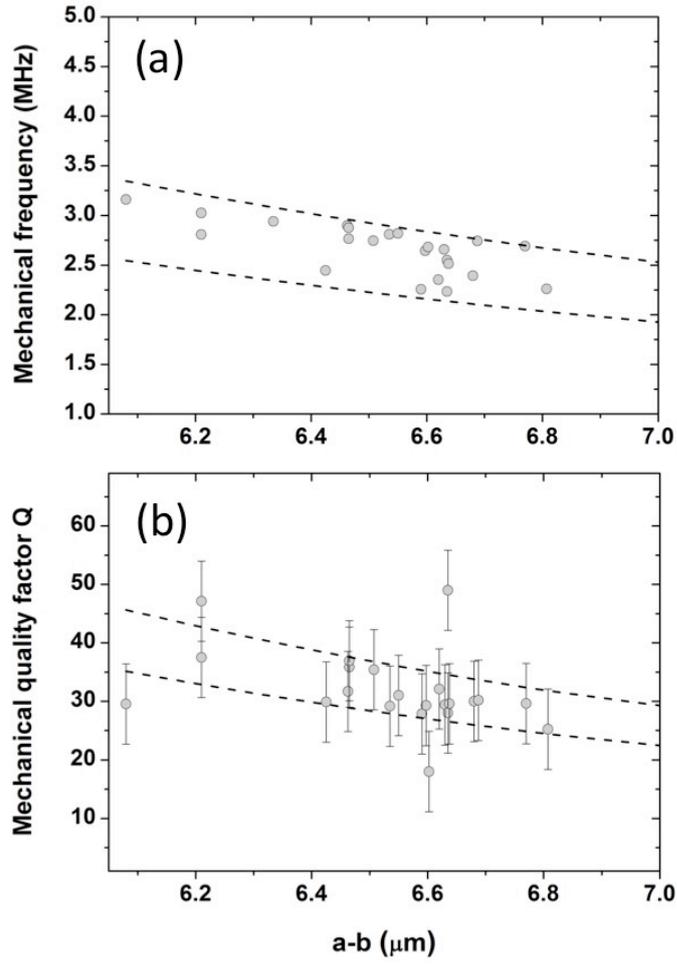

FIG. 3. a) Mechanical frequency of the (P=0,M=0) flexural mode as a function of the under-etch (*a-b*). Empty circles are measurements on a set of 24 disks. Dashed lines are two bounding theoretical predictions obtained from an effective elastic theory where the frequency scales with $1/(a-b)^2$. b) Mechanical Q factor of the (P=0,M=0) mode as a function of (*a-b*). Empty circles are measurements on 24 disks. The two dashed lines are obtained by combining analytical models for air damping and squeeze-film damping with the effective formulas for the mechanical frequency.